\documentclass[sn-vancouver, Numbered]{sn-jnl}%

\usepackage{xcolor}%
\usepackage{manyfoot}%

\usepackage{amsmath,amsfonts,amssymb}
\usepackage{graphicx}
\usepackage{subcaption}
\usepackage{placeins}
\usepackage{multirow}
\usepackage{tabularx}

\newcolumntype{q}{>{\hsize=.01\hsize}X}
\newcolumntype{s}{>{\hsize=.25\hsize}X}
\newcolumntype{t}{>{\hsize=.33\hsize}X}
\newcolumntype{h}{>{\hsize=.5\hsize}X}
\newcolumntype{z}{>{\hsize=.66\hsize}X}
\newcolumntype{L}{>{\hsize=\hsize \raggedright\arraybackslash}X}
\newcolumntype{R}{>{\hsize=\hsize \raggedleft\arraybackslash}X}
\newcolumntype{H}{>{\hsize=0.67\hsize \raggedleft\arraybackslash}X}
\newcolumntype{Y}{>{\centering\arraybackslash}X}

\usepackage{booktabs}
\usepackage{makecell}
 \usepackage{xfrac}
\usepackage{ulem}
\newcommand{\quotes}[1]{``#1''}

\begin{document}

\title[]{Assessment of the Axial Resolution of a Compact Gamma Camera with Coded Aperture Collimator}

\author*[1,2]{\fnm{Tobias} \sur{Mei{\ss}ner}}\email{publications@ibt.kit.edu}
\equalcont{These authors contributed equally to this work.}
\author[3,4]{\fnm{Laura Antonia} \sur{Cerbone}}\email{lauraantonia.cerbone-ssm@unina.it}
\equalcont{These authors contributed equally to this work.}
\author[4,5]{\fnm{Paolo} \sur{Russo}}\email{prusso@na.infn.it}
\author[1]{\fnm{Werner} \sur{Nahm}}\email{werner.nahm@kit.edu}
\author[2, 6, 7, 8]{\fnm{J{\"u}rgen} \sur{Hesser}}\email{juergen.hesser@medma.uni-heidelberg.de}

\affil[1]{\orgdiv{Institute of Biomedical Engineering (IBT)}, \orgname{Karlsruhe Institute of Technology (KIT)}, \orgaddress{\city{Karlsruhe}, \country{Germany}}}
\affil[2]{\orgdiv{Heidelberg University}, \orgname{Mannheim Institute for Intelligent Systems in Medicine (MIISM)}, \orgaddress{\city{Mannheim}, \country{Germany}}}
\affil[3]{\orgname{Scuola Superiore Meridionale}, \orgaddress{\city{Napoli}, \country{Italy}}}
\affil[4]{\orgdiv{INFN Sezione di Napoli}, \orgname{Istituto Nazionale di Fisica Nucleare}, \orgaddress{\city{Napoli}, \country{Italy}}}
\affil[5]{\orgdiv{Dipartimento di Fisica \quotes{Ettore Pancini}}, \orgname{Università di Napoli Federico II}, \orgaddress{\city{Napoli}, \country{Italy}}}
\affil[6]{\orgdiv{Interdisciplinary Center for Scientific Computing (IWR)}, \orgname{Heidelberg University}, \orgaddress{\city{Heidelberg}, \country{Germany}}}
\affil[7]{\orgdiv{Central Institute for Computer Engineering (ZITI)}, \orgname{Heidelberg University}, \orgaddress{\city{Heidelberg}, \country{Germany}}}
\affil[8]{\orgdiv{CZS Heidelberg Center for Model-Based AI}, \orgname{Heidelberg University}, \orgaddress{\city{Mannheim}, \country{Germany}}}

\abstract{
\textbf{Purpose:} Handheld gamma cameras with coded aperture collimators are under investigation for intraoperative imaging in nuclear medicine. Coded apertures are a promising collimation technique for applications such as lymph node localization due to their high sensitivity and the possibility of 3D imaging. We evaluated the axial resolution and computational performance of two reconstruction methods.\\
\textbf{Methods:} An experimental gamma camera was set up consisting of the pixelated semiconductor detector Timepix3 and MURA mask of rank $31$ with round holes of $0.08$\,mm in diameter in a $0.11$\,mm thick Tungsten sheet.
A set of measurements was taken where a point-like gamma source was placed centrally at $21$ different positions within the range of $12$ to $100$\,mm.
For each source position, the detector image was reconstructed in $0.5$\,mm steps around the true source position, resulting in an image stack. The axial resolution was assessed by the full width at half maximum (FWHM) of the contrast-to-noise ratio (CNR) profile along the z-axis of the stack. 
Two reconstruction methods were compared: MURA Decoding and a 3D maximum likelihood expectation maximization algorithm (3D-MLEM). \\ 
\textbf{Results:} While taking $4{,}400$ times longer in computation, 3D-MLEM yielded a smaller axial FWHM and a higher CNR. The axial resolution degraded from $5.3$\,mm and $1.8$\,mm at $12$\,mm to $42.2$\,mm and $13.5$\,mm at $100$\,mm for MURA Decoding and 3D-MLEM respectively. \\
\textbf{Conclusion:} Our results show that the coded aperture enables the depth estimation of single point-like sources in the near field. Here, 3D-MLEM offered a better axial resolution but was computationally much slower than MURA Decoding, whose reconstruction time is compatible with real-time imaging.}


\maketitle

\section{Background}
\label{sec:intro}
Accurate localization and comprehensible visualization of radioactive source distributions play a crucial role in nuclear medicine \cite{Farnworth2023, Peterson2011, Fujii2012}.
Gamma cameras are an established tool to image the distribution of gamma sources either as projection image or as tomography like in single photon emission tomography (SPECT)~\cite{Peterson2011}. Recently, compact gamma cameras are under investigation for intraoperative applications with different variants found~\cite{Farnworth2023, Ozkan2015, Kogler2020, Massari2016}.
The majority of those cameras use parallel or pinhole collimation to extract the spatial information of incoming gamma photons. Coded aperture imaging (CAI) was proposed as an alternative collimation technique, because it offers a better trade-off between resolution and photon harvesting~\cite{Fenimore1978}. By placing hundreds of small pinholes in a specific pattern on a radiopaque sheet, the directional information of gamma sources is encoded by the shadow of the mask cast on the detector. In order to obtain an interpretable image, reconstruction is required.

\begin{figure}[tb]
    \centering
    \includegraphics[width=0.6\textwidth, trim=0 0 5 6, clip]{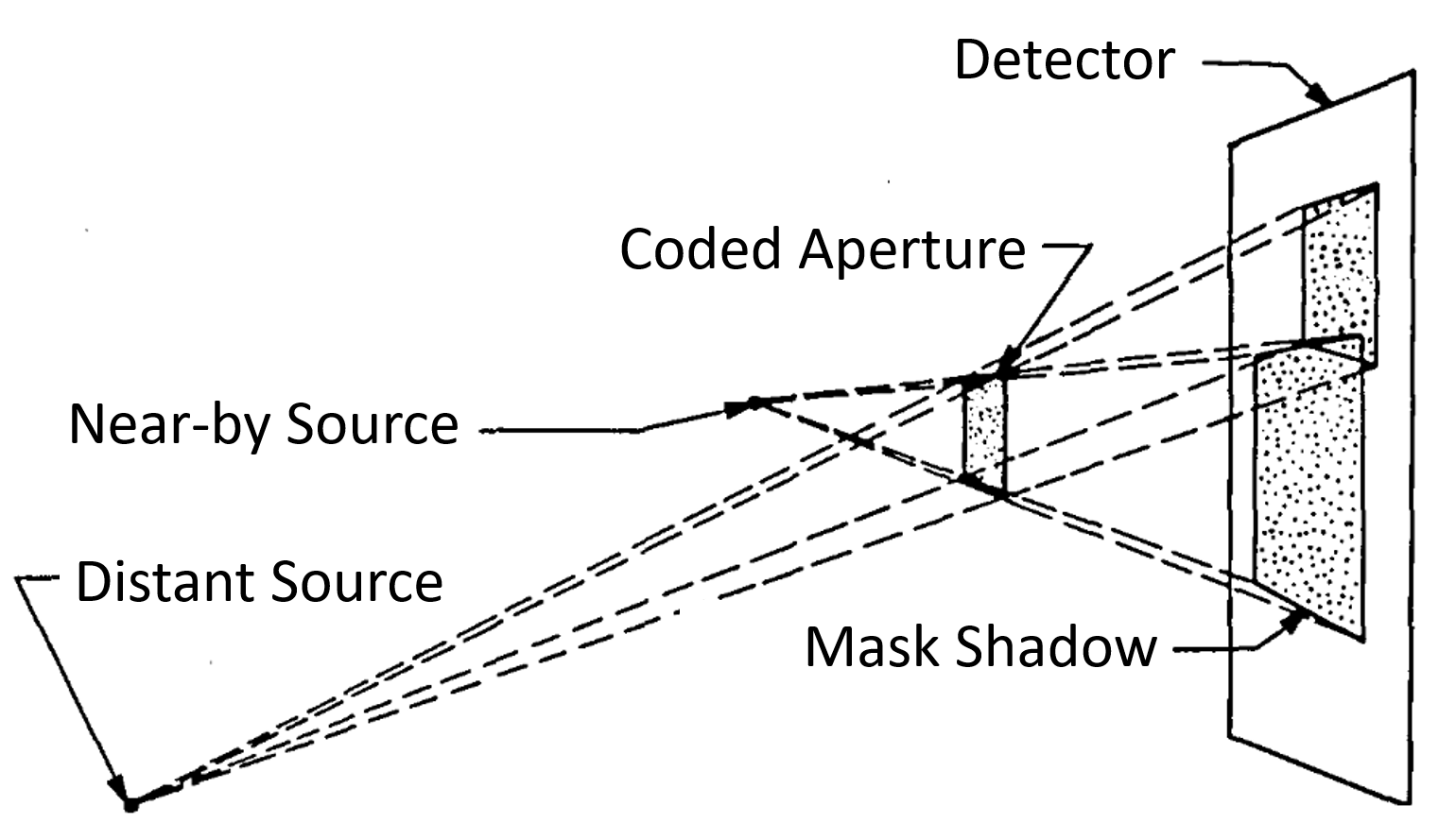}
    \caption{The basic principle of 3D coded aperture imaging (CAI): the lateral position is encoded by the shift of the shadow and the distance is encoded by the size of the shadow. Figure modified from~\cite{Cannon1979}.} 
    \label{fig:principle_3dcai}
\end{figure}

While other collimators do not require image reconstruction, for planar reconstruction in CAI an in-focus plane must be selected, i.e. a distance at which the source is assumed to be located. This, in principle, represents an inherent limitation of the approach but it enables obtaining a 3D reconstruction of the object from a single 2D detector image, at least for point sources~\cite{Cannon1979}:
by reconstructing the captured detector image at several subsequent planes, a 3D reconstruction of the emitting object can be obtained. 
As depicted in Fig.~\ref{fig:principle_3dcai}, the lateral position of a point source is encoded by the shift of the mask's shadow, while the source-to-mask distance is related to the size of the shadow.
The size of the shadow is the size of the coded aperture pattern that is projected on the detector with a magnification factor $M$ that can be described with the following relation~\cite{Cannon1979, Fujii2012}:

\begin{equation}
\label{eq:M}
    M = 1 +\,^b\!/\!_z
\end{equation}

\noindent with the detector-to-mask distance $b$ and the source-to-mask distance $z$.

Hence, a degradation in the axial resolution with growing source-to-mask distance can be expected.
Note that since the detector is encased within the camera body and thus not directly visible by the user, we chose this more intuitive definition, although others refer to the detector-to-source distance as $z$~\cite{Accorsi2001a, Mu2006}.

Multiple approaches to 3D imaging of gamma sources in the field of intraoperative surgery applications have been proposed: stereo gamma cameras~\cite{Kaissas2017, Paradiso2018}, simultaneously tracking and merging the image data of a freely movable camera~\cite{Kogler2020}, and direct 3D reconstruction from a single gamma camera with coded aperture~\cite{Fujii2012}.
The dilemma of the axial resolution of a single camera with a coded aperture is that the sensitivity is high when the camera is close to the source but at the same time, multiple near field effects deteriorate the quality of the reconstructed images~\cite{Accorsi2001a, Mu2006}.
To the best of the authors' knowledge, only two reconstruction algorithms have been developed and used for 3D CAI reconstruction: MURA Decoding and MLEM. The first retrieves a sequence of images at multiple distances from which the depth of a single source could be estimated by localizing the maximal response~\cite{Russo2020}. However, since the images reconstructed at different depths are independent of each other, being reconstructed as if the sources were all located at the same depth, this
blurs the object along the z-direction and provides a poor axial resolution.
The second is a 3D convolution-based maximum likelihood expectation maximization algorithm (MLEM) that was introduced to reconstruct an entire 3D source distribution~\cite{Mu2006, Mu2016}.

Localizing the 3D position of spherical gamma sources is important in sentinel lymph node (SLN) biopsy (SLNB), where radioactively marked lymph nodes in the axilla need to be found and dissected for breast cancer staging~\cite{Ibraheem2019, Farnworth2023}. Compact gamma cameras are conveniently used for SLNB~\cite{Ibraheem2019} for their good spatial resolution and sensitivity; we are here proposing a new compact gamma camera equipped with a coded aperture collimator, which could, in turn, increase the sensitivity by keeping an excellent spatial resolution and add the possibility of 3D imaging. 
SLNs are not point-like sources and can only be considered as such at larger distances, and it is known that extended sources are reconstructed with lower quality than point sources in CAI~\cite{Mu2006, Fujii2012}. 
Thus, the study presented in this paper with its point-like source provides a fundamental basis of the axial resolution in CAI and an in-depth analysis for extended sources is required.

In contrast to the extensive investigations into the lateral resolution of CAI~\cite{Accorsi2001a, Fujii2012, Rozhkov2020}, the axial resolution has received much less attention. Only a few articles exist, covering only a limited range of source-to-mask distances~\cite{Mu2006, Fujii2012, Russo2020}.
This paper aims at closing the existing gap and makes the following contributions:

\begin{enumerate}
    \item A systematic experiment and assessment of the axial resolution of a compact gamma camera equipped with a coded aperture collimator is presented.
    \item We propose a reproducible method for measuring the axial resolution by calculating the FWHM of the CNR profile along the z-axis of a point-like source.
    \item The 3D-MLEM algorithm from~\cite{Mu2006} is extended by a normalization factor and mask transmission that adapts the algorithm to a general camera setup. 
    \item This paper compares two coded aperture reconstruction methods and demonstrates that 3D-MLEM can be considered as the slower but superior reconstruction method compared to standard MURA Decoding which is faster but less precise.
\end{enumerate}

\noindent The entire acquired dataset of $21$ images and its preprocessed versions are publicly available at \url{https://zenodo.org/doi/10.5281/zenodo.8315861}.

\section{Material \& Methods}
\label{sec:mam}
This section describes our experimental gamma camera, how the image data were acquired, the reconstruction methods used and how the axial and lateral resolutions were determined. 

\subsection{Image Acquisition}
\label{sec:mm:imag_acq}
\begin{figure}[tb]
    \centering
    \includegraphics[width=1.0\textwidth, trim=0 0 0 0, clip]{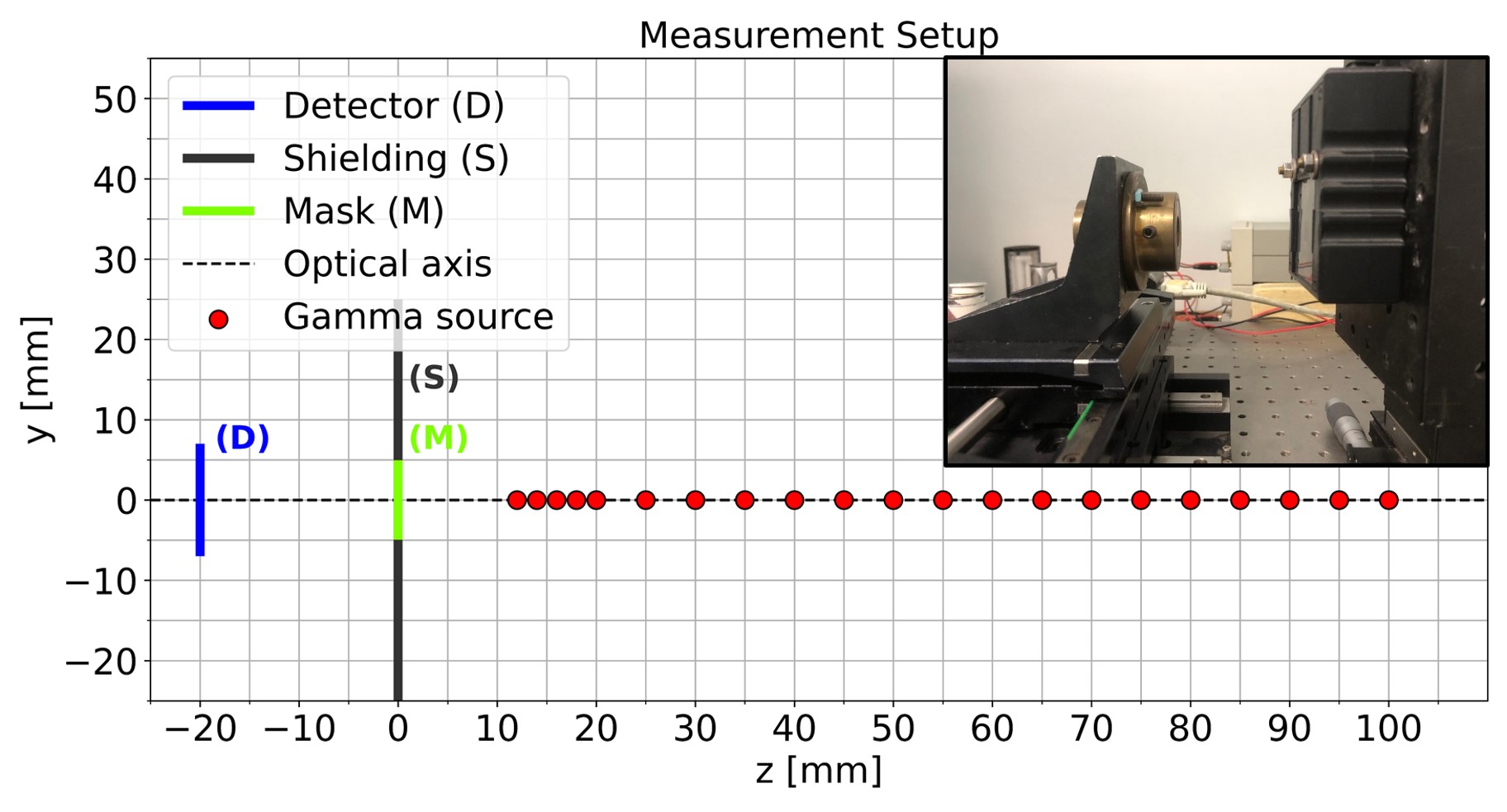}
    \caption{Images were captured with source-to-mask distances from $12$ to $20$\,mm in $2$\,mm steps and from $25$ to $100$\,mm in $5$\,mm steps (21 positions). The photograph in the top right corner shows the source holder on the left and the gamma camera in its housing with parts of the Tungsten mask visible on the right.} 
    \label{fig:img_acq}
\end{figure}

The experimental setup we used for image acquisition is composed of a Timepix3 application-specific integrated circuit (ASIC) bump bonded to a $0.5$\,mm thick Silicon sensor (MinipixEDU camera produced by Advacam) with a sensitive area of $14.08\times14.08$\,mm$^2$ coupled to a rank 31 no-two-holes-touching (NTHT) MURA Mask having $0.08$\,mm diameter holes in a $0.11$\,mm thick Tungsten sheet. The basic MURA pattern was duplicated in a $2\times2$ arrangement leading to a total mask size of $9.92\times9.92$\,mm$^2$.
This coded aperture collimator used for the compact gamma camera MediPROBE2 was found to provide the highest lateral resolution among the intraoperative gamma cameras currently available~\cite{Farnworth2023}.
We designed a 3D-printed case to keep the detector and the collimator in a fixed distance and axially aligned. The case holds the mask in a detector-to-mask distance $b$ of $20$\,mm. 
We used an automatic linear axis, which enabled us to move the source automatically and with high precision without interrupting the measurement procedure. An L-shaped holder was attached to the axis on which the $^\text{241}$Am source was clamped. The source's nominal diameter is $1$\,mm but was previously measured to have a FWHM of $0.65$\,mm~\cite{Bertolucci2002} and emits gamma photons of $59.5$\,keV. The source holder together with the gamma camera is depicted in Fig.~\ref{fig:img_acq}. The coded aperture mask was originally designed for sources of $30$\,keV; nevertheless, the source-mask combination used in this paper was the only one available to us. Besides, it has been shown that usage of this mask at higher energies ($80$\,kV X-ray beam) may reduce the image contrast but does not impede the imaging with a high CNR~\cite{Russo2011}.

The images were recorded with a software tool from Advacam called Pixet. Instead of collecting a predefined number of photons per measurement, we kept the acquisition time constant for each acquisition. We recorded 9{,}000 frames with an acquisition time of $0.1$\,s in \quotes{Tracking} mode. In this acquisition mode, the energy deposited by the interacting particles in the sensor is registered in each pixel, together with the time instant at which the interaction is revealed in the pixel. This information allows for the reconstruction of tracks released in the sensor from the impinging radiation, via a clustering algorithm based on time correspondence and spatial proximity of hits. This process, from which the name of the acquisition mode is derived, ultimately allows us to infer the type of radiation revealed by the detector based on its energy and the shape of its track. 
The time interval of $0.1$\,s per frame was chosen to avoid double counting of photons in a single pixel. A lower threshold of $5$\,keV was selected and no further energy windowing was applied (i.e. no hit was discarded from the final detector image based on the energy deposited in a pixel). Each acquisition lasted about $20$ minutes, including $15$ minutes active acquisition and time required for the intermediate processing of the detector. The pixel value in the resulting image represents the energy deposited in keV in each pixel integrated over the duration of the whole acquisition. 
We captured images at $21$ different source-to-mask distances in the range of $12$ to $20$\,mm in steps of $2$\,mm and from $20$ to $100$\,mm in $5$\,mm steps (Fig.~\ref{fig:img_acq}).

\paragraph{Preprocessing the detector images} 
To eliminate outliers and cope with erroneous pixels, the raw detector images were preprocessed before undergoing image reconstruction. Preprocessing contains outlier replacement by the median value of their $3\times3$ neighborhood. Per image, all pixels having values outside the range of the 1$^{\text{st}}$ and 99$^{\text{th}}$ percentile were considered to be outliers. Additionally, Gaussian smoothing with a sigma of $1$\,pixel was applied. Fig.~\ref{fig:mlem_stuff}b visualizes the raw and preprocessed data. The $21$ acquired detector images and their preprocessed versions are available at \url{https://zenodo.org/doi/10.5281/zenodo.8315861}.

\subsection{Reconstruction Methods}
\label{sec:mm:rec_meth}
We analyzed two different reconstruction methods for CAI: the most commonly used MURA Decoding and an improved version of 3D-MLEM.

\subsubsection{MURA Decoding}
\label{sec:mm:mura_dec}

The algorithm for MURA Decoding was already proposed with the invention of the uniformly redundant arrays (URA)~\cite{Fenimore1978}, the predecessor of today's widely used coded aperture pattern called Modified Uniformly Redundant Arrays (MURA)~\cite{Gottesman1989}. MURA Decoding is the inverse filter to the encoding pattern and is based on a linear and deterministic imaging model. 
The decoding pattern can directly be derived from the MURA pattern and its derivation from the encoding pattern can be found in~\cite{Cieslak2016}. 
The decoding pattern $d_z(x, y)$ used in the reconstruction step is magnified to target an in-focus plane according to the distance between the plane and the mask, denoted as $z$~\cite{Accorsi2008}. 
MURA Decoding makes use of circular convolution which is assured by two aspects: the $2\times2$ arrangement of the basic MURA pattern, and the use of the central part of the detector image $p(x, y)$. The central part, $\mathbf{C}\left(p(x, y), z\right)$, is the portion of the detector onto which one entire mask pattern (i.e. a quarter of the $2\times2$ arrangement) is projected. The size of this projection surely depends on the magnification factor $M$ given by Eq.~(\ref{eq:M}), therefore on $z$. As a result, the size - in pixels - of the reconstructed images depends on the source-to-mask distance. The reconstructed image $\hat{f}_{z}(x, y)$ is obtained by the following operation where \quotes{$\circledast$} denotes the circular 2D convolution operator:

\begin{equation}
\label{eq:mura_decoding}
    \hat{f}_{z}(x, y) = \mathbf{C}\left(p(x, y), z\right) \circledast d_z(x, y).
\end{equation}

\noindent MURA Decoding can be considered the most widely used reconstruction method in CAI, due to its simplicity and speed when carried out in the Fourier domain~\cite{Meißner2023}.
A limitation of MURA Decoding is the underlying linear and deterministic imaging model. Especially when imaging sources of low photon flux, degradation due to substantial Poisson noise contribution must be expected. Deviations from the linearity assumption are known and the induced systematic noise increases as the source moves closer to the camera~\cite{Mu2006, Accorsi2001}. The algorithm used in this paper has been implemented in MATLAB R2022b and to accelerate the process, the convolution is carried out in the Fourier domain.
The pixel values of the reconstructed images are of minor importance for this study and only the relative values affect the CNR and hence the axial resolution. 

\subsubsection{3D-MLEM}
\label{sec:3d_mlem}

\begin{figure}[tb]
    \centering
    \includegraphics[width=\textwidth, trim=3 10 3 5, clip]{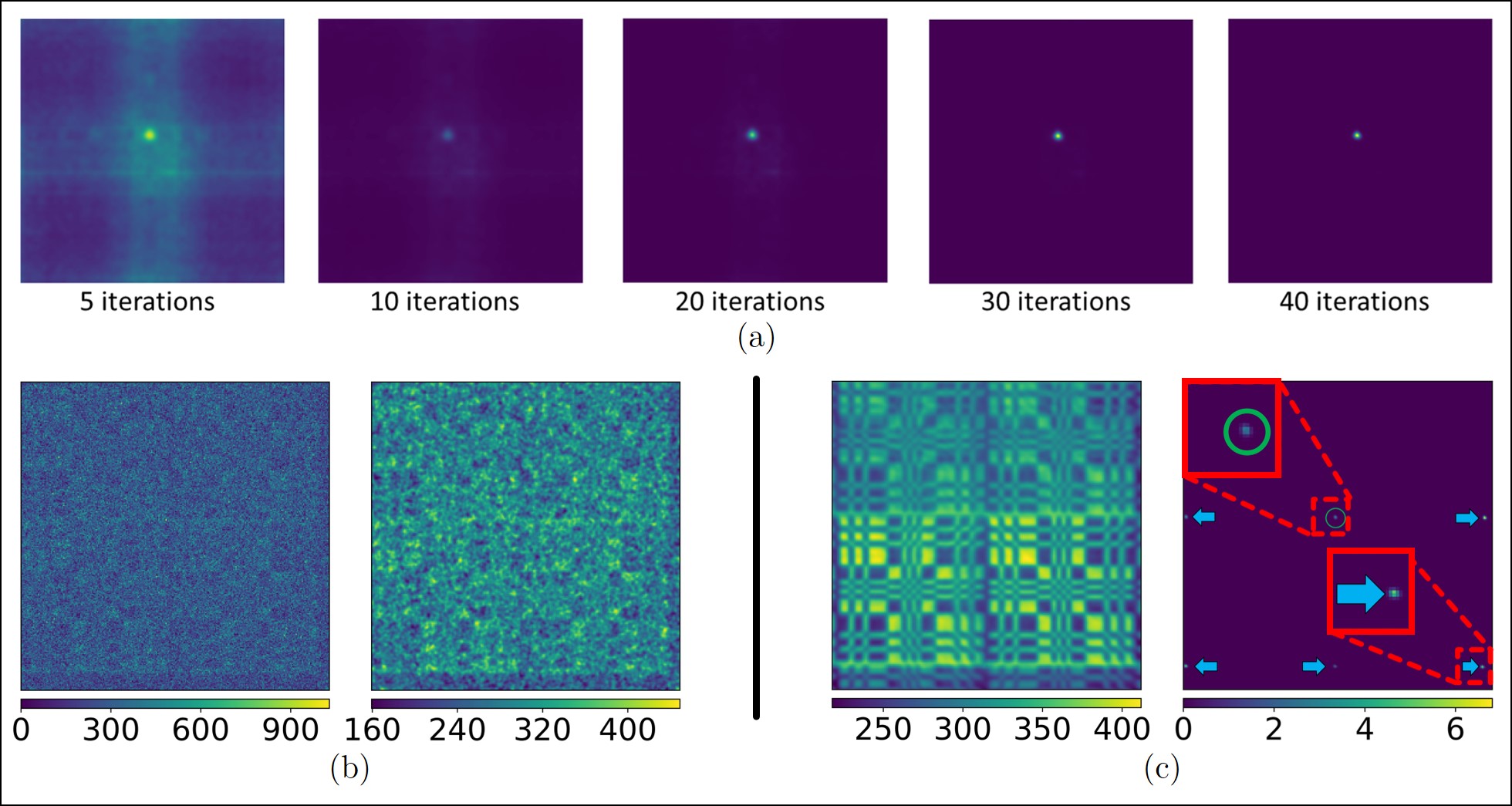}
    \caption{
    (a): Successive number of iterations for the 3D-MLEM algorithm applied to the source and reconstructed at $30$\,mm.
    (b): Raw (left) and preprocessed detector image (right). (c): The left image shows the forward projection of 3D-MLEM for the source at $40$\,mm and the right image shows the reconstruction at $40$\,mm. Due to the $2\times2$ arrangement of the basic MURA pattern, multiple ghost sources (blue arrows) along the image border emerge in addition to the true source (green circle) from the 3D-MLEM algorithm for sources that are more than $40$\,mm away.}    
    \label{fig:mlem_stuff}
\end{figure}

The maximum likelihood expectation maximization algorithm (MLEM) is an iterative algorithm that estimates the source distribution with the highest likelihood assuming the measured detected photons follow a random Poisson process~\cite{buzug2008algebraic}. The original MLEM algorithm was adapted to CAI by replacing the computationally expensive system matrix with a convolutional approach~\cite{Mu2006}. For more information, the reader is referred to~\cite{Meißner2023}. 
Additionally to the 2D reconstruction method, Mu et al.~\cite{Mu2006} extended this algorithm to differentiate between different source-to-mask distances. Thus, an entire 3D source distribution can be reconstructed. Similarly to MURA Decoding, the entire gamma camera is solely defined by its point-spread function (PSF) $h_z(x, y)$. The algorithm for the $(\text{k}+1)^{\text{th}}$ iteration of reconstructing the source distribution $\hat{f}_{z}^{(\text{k}+1)}(x, y)$ at a source-to-mask distance $z$ is given by

\begin{equation}
\label{eq:old_3d_mlem}
    \hat{f}^{(\text{k}+1)}_z = \frac{\hat{f}^{(\text{k})}_z}{\sum_{z}^{} h_z} \bullet 
    \left[
    h_z
    \times 
    \frac
    {
    p - 
    \displaystyle\sum_{z' \neq z}^{} \hat{f}^{(\text{k})}_{z'} \ast h_{z'}
    }
    {
    \hat{f}_z^{(\text{k})} \ast h_z
    }
    \right] 
\end{equation}

\noindent where \quotes{$\ast$} represents the linear 2D convolution, \quotes{$\times$} the 2D correlation and \quotes{$\bullet$} the point-wise multiplication. For better readability the lateral coordinates $(x,y)$ have been omitted.

\noindent However, two aspects are not taken into account by 3D-MLEM: first, a mask that is smaller in dimension than the detector and second, transmission of gamma photons through the radiopaque material. As these two conditions were verified in our experimental setup, we extended the 3D-MLEM algorithm with two modifications: a more general normalization term $n_z$ that accounts for the size difference between the mask and the detector and a forward simulation function $\mathbf{F}$, which incorporates mask transmission.

\paragraph{Normalization term}
The 3D-MLEM formula is derived from the general MLEM formula with the system matrix $\textbf{A}$, where its entries $a_{ij}$ denote the probability that photons from source $j$ are detected in detector pixel $i$. In literature the following normalization term, also referred to as sensitivity, can be found~\cite{buzug2008algebraic}:

\begin{equation}
    n_j = \displaystyle\sum_{i=0}^{n} \textbf{A}_{ij}
\end{equation}

\noindent This summation over the matrix columns represent the summed likelihood that the photon from source $j$ is detected by any pixel of the detector.
$n_j$ is a function over $j$ and in general, this is not a homogeneous distribution. When the coded aperture mask is smaller than the detector, photons that are emitted further outside the center have a higher possibility to pass through a pinhole but not hitting the detector. One can imagine this by looking at the mask shadow cast by an off-center point source, where only part of the shadow hits the detector. The further off-center the source is, the higher the share of the shadow that is not falling within the detector area and the smaller the summed likelihood $n_j$ becomes.
The normalization factor $n_{j}$ ensures that the inherent forward projection has approximately as many detected photons as the given detector image $p$.
\noindent Translated to the convolution-based 3D-MLEM algorithm the normalization factor $n_j$ is an image that depends on the source-to-mask distance $n_z$. It can be calculated offline by a backward projection of an entirely illuminated detector to the plane in focus. Thus, we obtain $n_z$ by calculating the cross-correlation between an all $1$ image ($\mathbf{1}$) and the PSF at the given distance $z$: 

\begin{equation}
    n_z = h_z \times \mathbf{1}
\end{equation}

\paragraph{Forward simulation with transmission}
Transmission noise emerges from photons that penetrate the mask and in this paper it is approximated as uniform background noise proportional to the transmission coefficient of the coded aperture mask. With a mask thickness of $0.11$\,mm and a $59.5$\,keV source (see Sec.~\ref{sec:mm:imag_acq}) the transmission probability $t$ of our setup is approximately $46$\%, meaning that about half of the photons pass through the mask. Because this transmission is high, the forward projection $\mathbf{F}$ of the 3D-MLEM algorithm had to be adapted. The projected image becomes a weighted superposition of the projection of the reconstruction $f_z$ and a uniform transmission noise image. The weight is set according to the transmission rate $t$ and the sum of emitted photons from the in-focus plane $f_z$:

\begin{equation}
    \label{eq:forward_sim}
    \mathbf{F}\left( f_z, h_z\right) = 
    (1-t)  \left( f_z \ast h_z \right) + t\, \sum_{x, y} f_z .
\end{equation}

\noindent Additionally, we do not use the PSF of the NTHT pattern but of the two-holes-touching (THT) pattern to avoid periodical background noise~\cite{Meißner2023}. \\
In contrast to MURA Decoding, we decided to use the entire detector image for the reconstruction. This choice ensures that the output maintains a consistent size of $256\times256$ pixels. Because MLEM is based on the convolutional model, the field of view (FOV) is equal to the FOV of a single pinhole collimator.
All in all, the proposed 3D-MLEM algorithm can be summarized as follows:

\begin{equation}
\label{eq:new_3d_mlem}
    \hat{f}^{(\text{k}+1)}_z = \frac{\hat{f}^{(\text{k})}_z}{n_z} \bullet 
    \left[
    h_z
    \times 
    \frac
    {
    p - 
    \displaystyle\sum_{z' \neq z}^{} \mathbf{F}\left(\hat{f}^{(\text{k})}_{z'}, h_{z'}\right)
    }
    {
    \mathbf{F}\left( \hat{f}_z^{(\text{k})} \ast h_z \right)
    }
    \right] .
\end{equation}

The 3D-MLEM algorithm assumes that the pixel intensity represents photon hits. Our detector images, however, represent deposited energy per pixel. The conversion to photon hits is not trivial due to charge sharing between neighboring pixels. Comparing the axial resolution of the source at $50$\,mm based on the detector image and on the same image divided by $59.5$\,keV (ignoring any charge sharing), the absolute difference was less than $2.28$\,µm. Thus, we decided to not use any conversion and directly apply the 3D-MLEM to our captured detector images. As with MURA Decoding, the pixel values of the reconstructed images do not affect the axial resolution and only the relative intensities will influence the CNR.
We decided to apply 40 iterations for the reconstruction from the acquired detector images (see Fig.~\ref{fig:mlem_stuff}a) since, from visual inspections, we found that to be a good compromise between noise amplification and reconstruction quality.
The 3D-MLEM algorithm was implemented in Python (3.8) using the NumPy (1.24) library and, similar to MURA Decoding, all convolutions are performed in the Fourier domain.

\subsection{Assessing the axial resolution}
\label{sec:mm:ax_res}
The axial resolution expresses how well a point-like source can be localized in the depth direction, i.e. along the z-axis. 
We use the profile of the contrast-to-noise ratio (CNR) along the z-direction to determine the axial resolution as the full width at half maximum (FWHM) of this Gaussian-like curve: this provides a more intuitive understanding of the spatial resolution and takes into account not just the source intensity - as in~\cite{Fujii2012}, where the pixel intensity of the source was used to compute the axial spatial resolution - but also how well the source can be distinguished from the background noise.
The following definition of the CNR was employed:
\begin{equation}
\label{eq:cnr}
    CNR = \frac{\bar{S} - \bar{B}}{\sigma_{B}}\,, 
\end{equation}
\noindent where $\bar{S}$ denotes the mean intensity of the signal near the true source position, while $\bar{B}$ and $\sigma_{B}$ are respectively the mean intensity and the standard deviation of the background.

First, we reconstructed each source's image within a broad range from $5$ to $100$\,mm in $5$\,mm steps to locate the source. Second, we reconstructed images within a tighter $z$ range containing the actual source position in $0.5$\,mm steps resulting in sets of images ranging from 60 to 240 images for MURA Decoding and 54 to 101 for 3D-MLEM.
From here on, we will refer to a set of reconstructions of the same detector image at different depths as an \textit{image stack}. For easier handling, the reconstructions from MURA Decoding were resized by bilinear interpolation to the image size of the reconstruction at the true source position. %
To quantify the impact of the resizing procedure, the axial resolution was computed for both the resized and non-resized stacks of reconstructions for a source placed at $30$\,mm. The resulting values were $11.9 \pm 0.5$\,mm for the resized and $12.5 \pm 0.5$\,mm for the non-resized one, which are compatible within the errors.
Therefore, for the sake of simplicity, we decided to continue our analysis on the resized stacks. No resizing was required for 3D-MLEM images, as the algorithm returns images of a fixed size.

To determine the CNR for each image of a stack, the signal $S$ and background $B$ is required. To minimize the influence of the operator on the CNR computation, we wrote a semi-automatic algorithm that sampled the whole image, thus avoiding manually choosing regions of interest (ROIs) for $S$ and $B$ and ensuring the reproducibility of the procedure. The diameter of the ROI was chosen separately for each stack, i.e. each captured detector image, based on the true source size (see Sec.~\ref{sec:mm:imag_acq}) and the true source distance. First, the FWHM source diameter of $0.65$\,mm was converted to pixels with respect to the FOV at the true source distance and then it was rounded to the nearest integer number to obtain the ROI diameter. The equation for the FOV of MURA Decoding was taken from~\cite{Fujii2012} and for 3D-MLEM the FOV of a pinhole collimator is used.

An \texttt{imageJ} macro sampled all possible positions of the ROIs per image stack: depending on the image size and ROI diameter there were between approximately $14{,}000$ and $31{,}000$, and $54{,}000$ and $65{,}000$ ROI positions per image for MURA Decoding and 3D-MLEM.
For each ROI its position in pixels, the average intensity, and standard deviation were calculated and stored for each image of the stack. 
The ROI with the highest average intensity in the in-focus image (i.e. the image reconstructed at the $z$ where the source was actually located) was selected as the signal $S$. A constraint was introduced that restricted the signal ROI to be in the inner $50$\% of the image area to avoid measuring one of the \textit{ghost sources} along the image border as visible in Fig.~\ref{fig:mlem_stuff}c.
Furthermore, ROIs that overlapped with $S$ were discarded from further processing. All other ROIs were selected as background ROIs and the background mean $\bar{B}$ was computed as the average intensity of all background ROIs; the same was performed for the standard deviation: $\bar{\sigma_B}$.
Once the signal ROI was found in the in-focus image its position was kept the same for each image of the stack. The same was done for the ROIs where $\bar{B}$ and $\bar{\sigma_B}$ were computed. 

Finally, a Gaussian curve with offset of the following form was fitted through the CNR profile: $\text{CNR}(z) = \alpha +(\beta-\alpha)\,\exp\left(-(z - \gamma)^2 / (2 \, \delta^2)\right)$ with the fitting parameters $\alpha$, $\beta$, $\gamma$, and $\delta$. The fitting procedure was carried out in Python (3.8) with the \texttt{curve\_fit} function from SciPy (1.10). The axial resolution and its standard deviation through the fitting procedure are reported as FWHM with following correspondence between the Gaussian's standard deviation $\delta$ and the FWHM~\cite{Russo2020}: 

\begin{equation}
    \text{FWHM} = 2\sqrt{2\ln{2}} \,\,\delta \approx 2.35 \,\, \delta
\end{equation}

\subsection{Assessing the lateral resolution}
\label{sec:mm:lat_res}
In order to compare our results with values from literature we measured the lateral resolution as well. This allows us to additionally report the axial resolution relative to the lateral resolution. Following the suggestion given in a recent review on intraoperative gamma cameras~\cite{Farnworth2023}, we determined the lateral resolution for $30$\,mm, $50$\,mm, and $100$\,mm source-to-mask distance. 

To compute the lateral resolution, we first selected the in-focus image within the image stack used for the assessment of the axial resolution. We then took the source profile along the row with the highest pixel intensity. A Gaussian curve with offset was fitted to this profile and the FWHM value was obtained from the resulting standard deviation. As the FWHM value was given in pixels, it was converted to mm by using the FOV of the respective reconstruction method.

\section{Results}
\label{sec:res}

\subsection{Proposed 3D-MLEM algorithm}
\label{sec:res:new_mlem}
\begin{figure}[ht]
    \centering
    \includegraphics[width=1\textwidth, trim=5 5 5 5, clip]{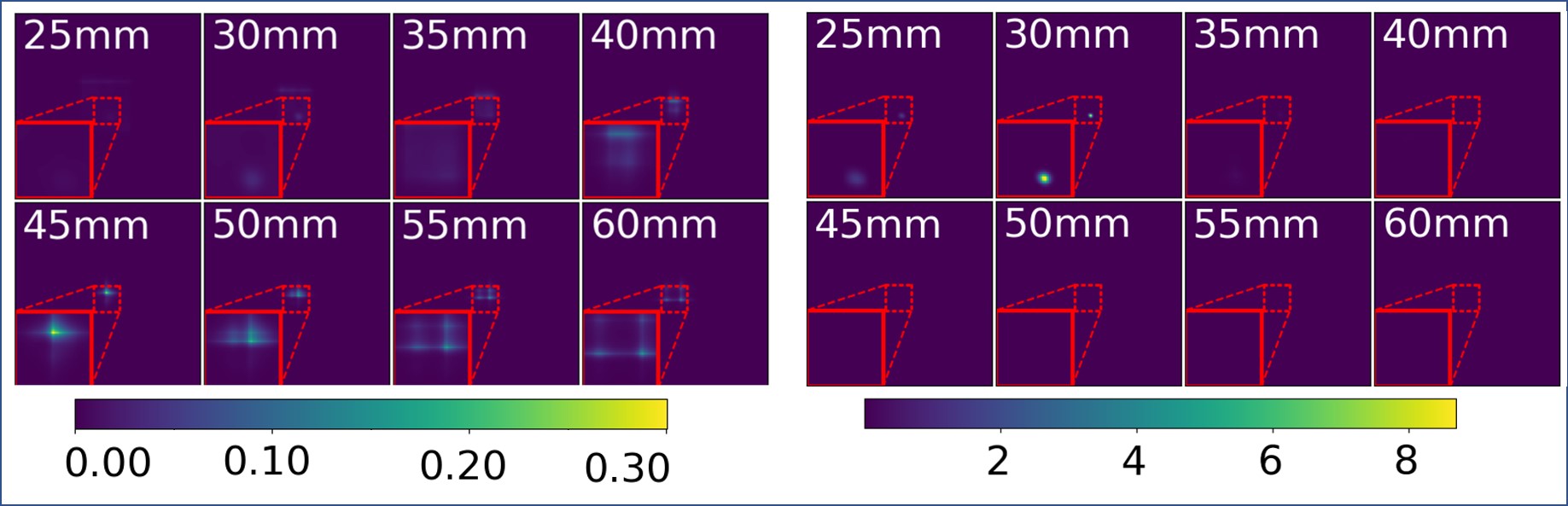}
    \caption{The proposed 3D-MLEM algorithm (right) in comparison to the original 3D-MLEM (left) from~\cite{Mu2006} applied to the source at $30$\,mm distance. The center, marked by the red square, has been magnified for better visualization.}
    \label{fig:old_vs_new_3dmlem}
\end{figure}

A comparison between the initial 3D-MLEM and 3D-MLEM with our proposed modifications is shown in Fig.~\ref{fig:old_vs_new_3dmlem}. Note how the source at $30$\,mm distance is barely visible in the initial 3D-MLEM reconstruction. Additionally, the maximum intensity of the image stack is located in the reconstructed image at $45$\,mm. The same artifact is also visible in all other image stacks of sources at different distances. The image stack of our proposed 3D-MLEM algorithm shows a single bright spot at a distance where we indeed placed the source, despite the high transmission noise. 

\subsection{3D reconstructions}
\label{sec:res:3d_rec}
All 21 detector images were reconstructed within a range from $5$ to $100$\,mm in $5$\,mm steps to roughly locate the source in the axial direction. The first eight of the 20 images from the image stack of the source at $30$\,mm distance are shown in Fig.~\ref{fig:3d_comp_30mm}.
The entire image stack for the sources at $50$ and $100$\,mm can be found in Appendix~\ref{app:50and100}. 
Both methods show bright spots in the center of the reconstructed image at the true distance. 
Overall, the background of the 3D-MLEM images looks more uniform while MURA Decoding yields images with a higher background noise. For reconstructions at $50$\,mm (pixel intensity normalized to the range of $0$ to $1$) with the source positioned at $z=50$\,mm from the collimator, we yield a $\sigma_B$ of $0.0281$ and of $8.7614\cdot 10^-5$ for MURA Decoding and 3D-MLEM. From Fig.~\ref{fig:3d_comp_30mm} itself a worse axial resolution can already be seen for MURA Decoding, since the source is also visible in the reconstructions at $25$ and $35$\,mm distance. Additionally, 3D-MLEM is capable of reconstructing at distances that are closer than $15$\,mm.

\begin{figure}[tb]
    \centering
    \includegraphics[width=1.0\textwidth, trim=5 5 5 5, clip]{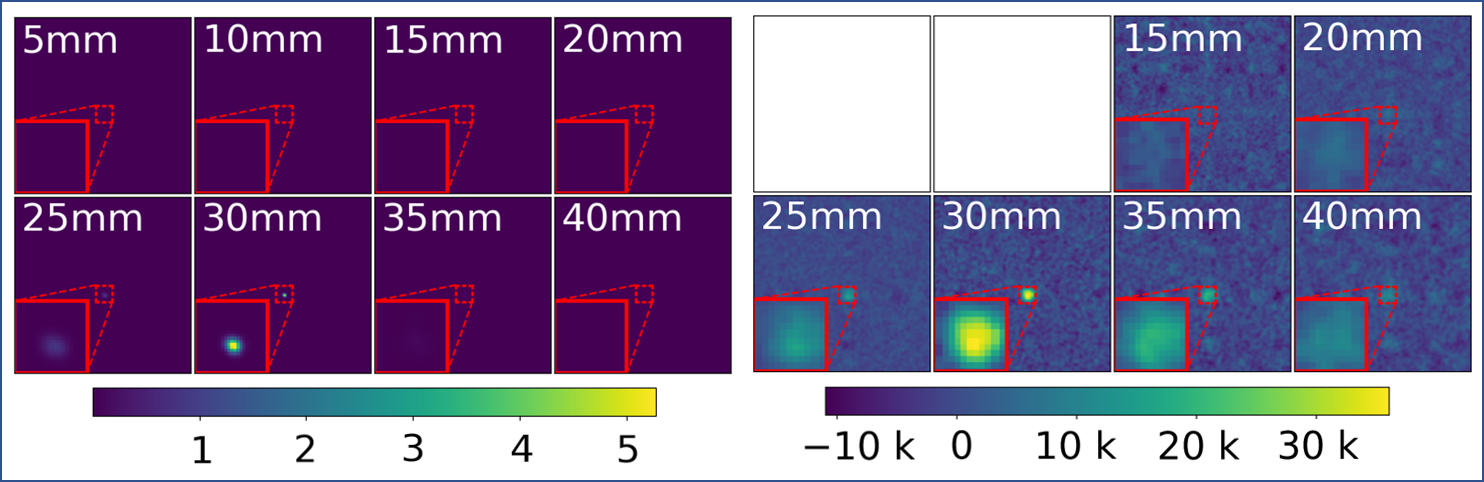}
    \caption{3D-MLEM (left) and MURA Decoding (right) of the $30$\,mm source. The distance between the mask and the in-focus plane in mm is indicated in the top left corner. MURA Decoding is not capable of reconstructing planes that are closer than $11$\,mm.
    A magnification of the area around the source (red dotted square) is shown in the bottom left corner.}
    \label{fig:3d_comp_30mm}
\end{figure}

\subsection{Axial resolution}
\label{sec:res:ax_res}

\begin{table}[htbp]
\centering
\caption{The FWHM axial resolutions are displayed separately for the two reconstruction methods (MURA Decoding and 3D-MLEM) and for raw and preprocessed detector images. The standard deviation values are obtained through the fitting algorithm.}
\label{tab:ax_res}
\begin{tabularx}{1.0\linewidth}{HRRRR}
\toprule
\textbf{Source} & \multicolumn{2}{c}{\textbf{MURA Decoding}} & \multicolumn{2}{c}{\textbf{3D-MLEM}} \\
\textbf{distance} & \multicolumn{1}{r}{raw} & \multicolumn{1}{r}{preprocessed} & \multicolumn{1}{r}{raw} & \multicolumn{1}{r}{preprocessed} \\
\text{[mm]} & \text{[mm]} & \text{[mm]} & \text{[mm]} & \text{[mm]} \\
\midrule
12  & 5.3 $\pm$ 0.6  & 5.3 $\pm$ 0.6  & 2.17 $\pm$ 0.03            & 1.75 $\pm$ 0.03 \\
14  & 4.7 $\pm$ 0.1  & 4.7 $\pm$ 0.1  & 2.20 $\pm$ 0.02            & 1.80 $\pm$ 0.06 \\
16  & 6.0 $\pm$ 0.2  & 5.8 $\pm$ 0.2  & 2.32 $\pm$ 0.04            & 1.85 $\pm$ 0.08 \\
18  & 6.9 $\pm$ 0.2  & 6.2 $\pm$ 0.2  & 2.34 $\pm$ 0.02            & 2.02 $\pm$ 0.03 \\
20  & 7.5 $\pm$ 0.1  & 7.3 $\pm$ 0.1  & 2.60 $\pm$ 0.09            & 2.26 $\pm$ 0.02 \\
25  & 11.0  $\pm$ 0.4 & 9.8 $\pm$ 0.4 & 2.37 $\pm$ 0.02            & 2.54 $\pm$ 0.04 \\
30  & 12.1  $\pm$ 0.5 & 11.9 $\pm$ 0.5 & 2.74 $\pm$ 0.02            & 2.76 $\pm$ 0.11 \\
35  & 17.2  $\pm$ 0.9 & 15.1 $\pm$ 0.8 & 2.45 $\pm$ 0.10            & 2.01 $\pm$ 0.14 \\
40  & 16.9  $\pm$ 0.9 & 18.5 $\pm$ 1.1 & $\dagger$4.32 $\pm$ 0.12   & $\dagger$3.51 $\pm$ 0.15 \\
45  & 14.2  $\pm$ 0.8 & 18.4 $\pm$ 1.3 & $\dagger$5.48 $\pm$ 0.06   & $\dagger$4.69 $\pm$ 0.13 \\
50  & 15.6  $\pm$ 1.0 & 17.5 $\pm$ 1.0 & $\dagger$5.53 $\pm$ 0.09   & $\dagger$5.97 $\pm$ 0.09 \\
55  & 18.3  $\pm$ 1.1 & 18.8 $\pm$ 1.1 & $\dagger$5.52 $\pm$ 0.11   & $\dagger$4.73 $\pm$ 0.08 \\
60  & 19.4  $\pm$ 0.9 & 19.9 $\pm$ 0.9 & $\dagger$6.71 $\pm$ 0.11   & $\dagger$5.24 $\pm$ 0.10 \\
65  & 23.9  $\pm$ 0.9 & 28.0 $\pm$ 1.2 & $\dagger$7.04 $\pm$ 0.12   & $\dagger$6.67 $\pm$ 0.10 \\
70  & 22.2  $\pm$ 0.7 & 23.8 $\pm$ 0.9 & $\dagger$8.19 $\pm$ 0.17   & $\dagger$7.37 $\pm$ 0.11 \\
75  & 26.6  $\pm$ 0.7 & 27.8 $\pm$ 0.7 & $\dagger$9.79 $\pm$ 0.17   & $\dagger$9.10 $\pm$ 0.17 \\
80  & 32.6  $\pm$ 0.5 & 35.4 $\pm$ 0.5 & $\dagger$10.40 $\pm$ 0.18   & $\dagger$11.64 $\pm$ 0.27 \\
85  & 32.7  $\pm$ 0.2 & 35.9 $\pm$ 0.4 & $\dagger$11.11 $\pm$ 0.21  & $\dagger$12.34 $\pm$ 0.24 \\
90  & 37.0  $\pm$ 0.4 & 37.8 $\pm$ 0.5 & $\dagger$11.20 $\pm$ 0.24   & $\dagger$10.37 $\pm$ 0.24 \\
95  & 35.2  $\pm$ 0.8 & 38.3 $\pm$ 0.8 & $\dagger$12.28 $\pm$ 0.29  & $\dagger$14.81 $\pm$ 0.41 \\
100 & 34.8  $\pm$ 0.4 & 42.2 $\pm$ 0.9 & $\dagger$14.84 $\pm$ 0.66  & $\dagger$13.48 $\pm$ 0.86 \\
\bottomrule
\end{tabularx}
\begin{tablenotes}
  \scriptsize
  \item[$\dagger$] Additional ghost sources appear in the resulting image stack.
\end{tablenotes}
\end{table}

\begin{figure}[tb]
    \centering
    \includegraphics[width=1.0\linewidth, trim=6 6 4 4, clip]{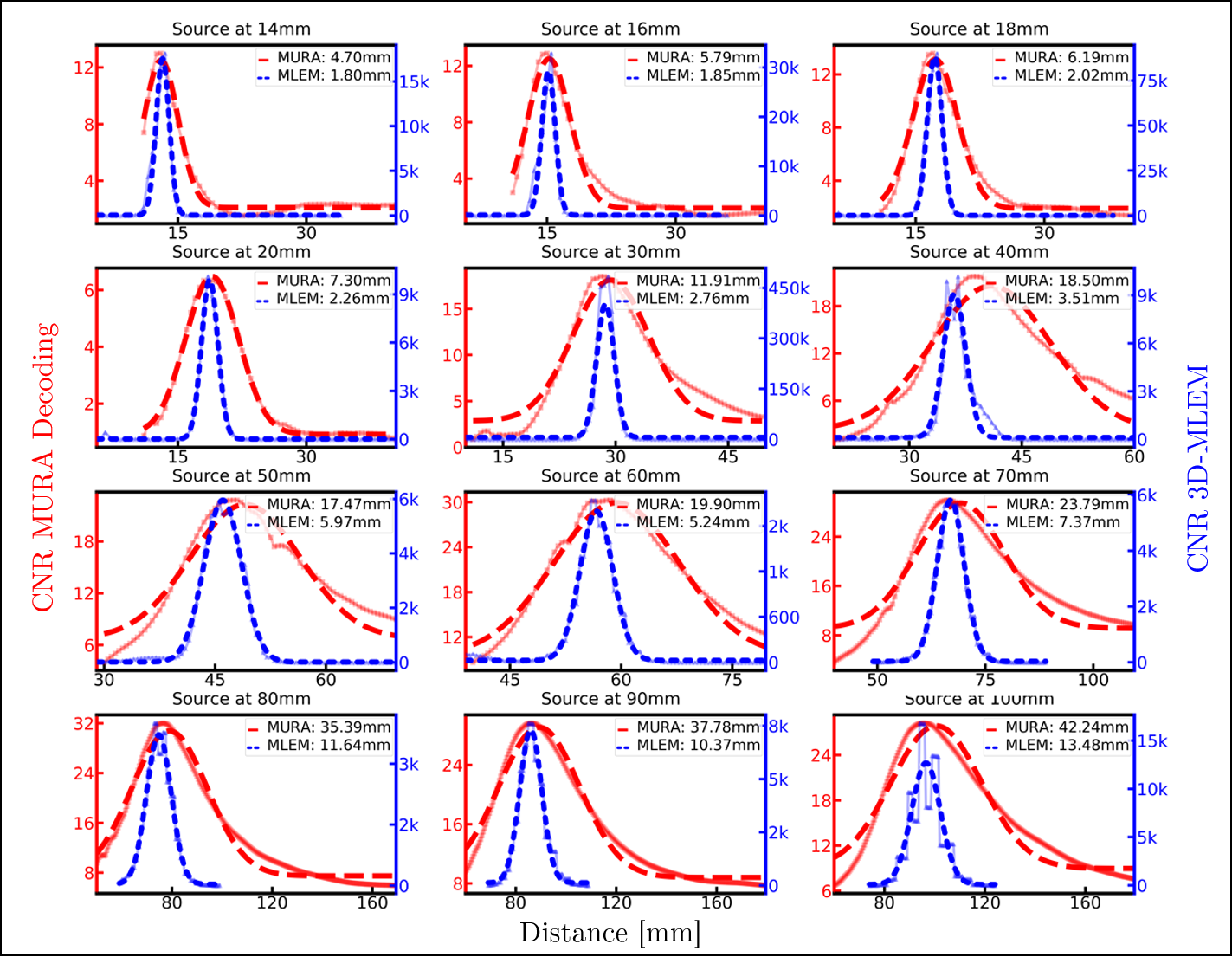}
    \caption{The CNR profiles over the distance used for reconstruction for a selection of source positions: the semi-transparent red line of squares and the blue line of triangles show the CNR profiles of MURA Decoding and 3D-MLEM reconstruction. The Gaussian curves with offset, represented by the bold red dotted line (MURA Decoding) and the blue dashed line (3D-MLEM), were fitted to the CNR profiles. These curves serve as the basis for determining the axial resolution, and the corresponding FWHM values are displayed in the top right corner of each graph.}    
    \label{fig:multi_ax_res}
\end{figure}

\begin{figure}[tb]
    \centering
    \includegraphics[width=1.0\textwidth, trim=0 10 0 0, clip]{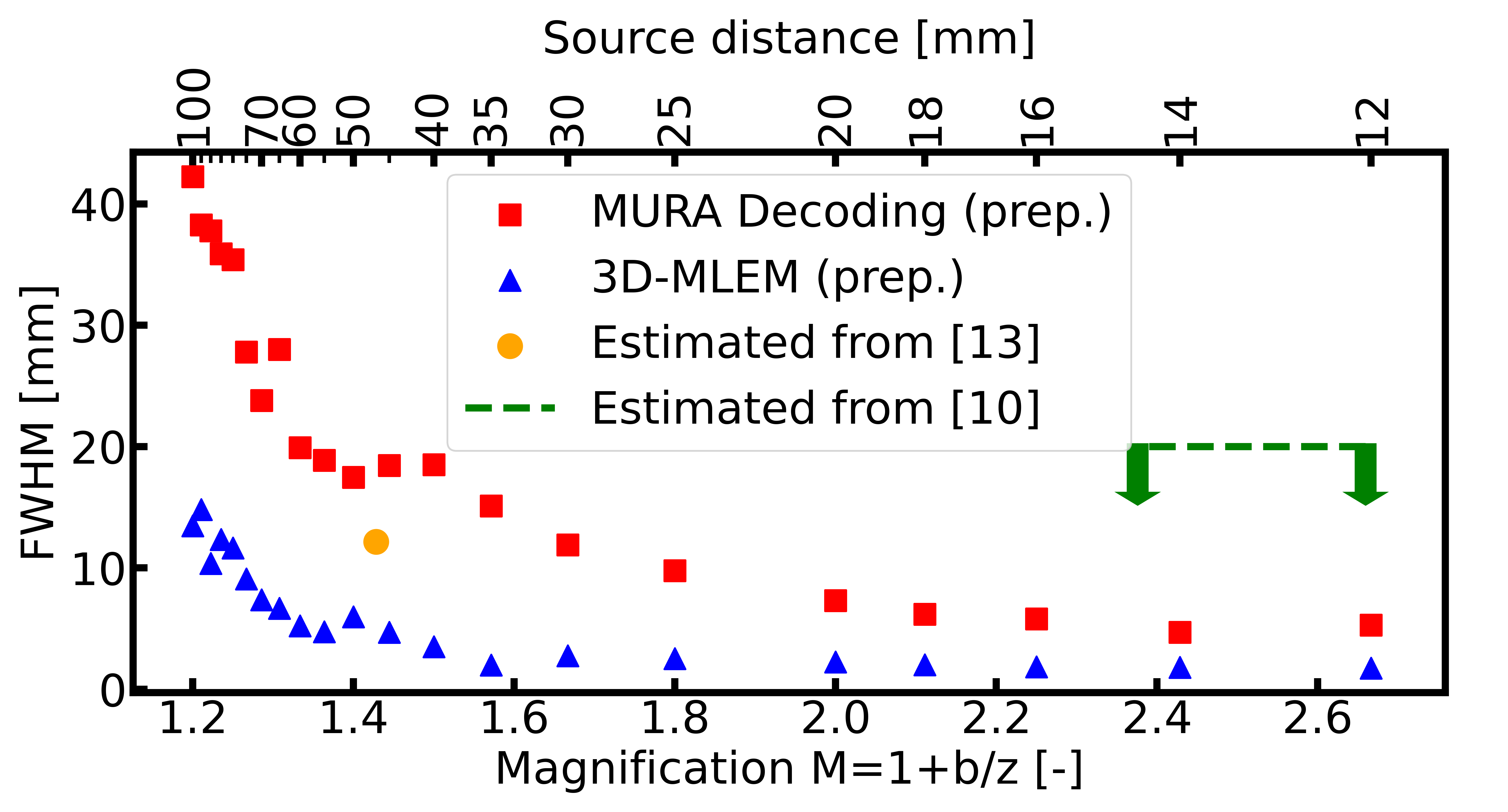}
    \caption{The axial resolution for both presented reconstruction methods plotted against the dimensionless magnification factor $M$. The orange circle and green dashed line represent reference values for the axial resolution estimated from literature~\cite{Russo2020, Mu2006}. Note that the source-to-mask distance at the top axis only corresponds to values from this paper.}    
    \label{fig:axial_res_plot}
\end{figure}

Table~\ref{tab:ax_res} shows the resulting FWHM axial resolution for each of the captured 21 images. Fig.~\ref{fig:multi_ax_res} presents the CNR profiles and Gaussian fits of a few image examples, from which the FWHM was derived. A clear difference between MURA Decoding and 3D-MLEM reconstruction can be seen: first, the axial resolutions obtained by 3D-MLEM are better (smaller in value), as their profiles are narrower than those obtained with MURA Decoding and, additionally, the CNR values, i.e. the height of the Gaussian curves, are greater by a factor between approximately $60$ and $30{,}000$ for the 3D-MLEM. 
The axial resolutions between raw and preprocessed detector images are almost identical. On average the ratios of the FWHM from preprocessed to raw detector images are $1.05\pm 0.10$ (MURA Decoding) and $0.93 \pm 0.12$ (3D-MLEM). However, the final objective of CAI reconstruction is to obtain a clear and interpretable image, and Fig.~\ref{fig:recos_prep_vs_raw} from the Appendix~\ref{app:raw_vs_prep_z20@20} shows that reconstructions based on the preprocessed images contain less background noise.
Therefore, from here on the axial resolution of the preprocessed images will be discussed.
The axial resolutions are $11.9 \pm 0.5$\,mm and $2.76 \pm 0.11$\,mm (source at $30$\,mm, magnification $M=1.67$), $17.5 \pm 1.0$\,mm and $5.97 \pm 0.09$\,mm (source at $50$\,mm, $M=1.40$), and $42.2 \pm 0.9$\,mm and $13.48 \pm 0.86$\,mm (source at $100$\,mm, $M=1.20$) for MURA Decoding and 3D-MLEM respectively. The average standard deviations introduced by the fitting procedure are $3.9$\,\% (MURA Decoding) and $2.7$\,\% (3D-MLEM). For easy comparison with other imaging systems, the axial resolution is plotted in Fig.~\ref{fig:axial_res_plot} against the magnification factor $M$, with the corresponding source-to-mask distance values. For sources at a distance greater than $40$\,mm, 3D-MLEM reconstructs up to eight \textit{ghost sources} in a regular pattern surrounding the true central position, as shown in Fig.~\ref{fig:mlem_stuff}c. 

\subsection{Lateral resolution}
\label{sec:res:lat_res}
The lateral resolutions based on the preprocessed detector images measured for MURA Decoding and 3D-MLEM are respectively $0.74$\,mm and $0.27$\,mm at a source distance of $30$\,mm, $0.80$\,mm and $0.29$\,mm  at $50$\,mm and $1.04$\,mm  and $0.4$\,mm at $100$\,mm. This results in ratios between axial and lateral resolutions of $16\!:\!1$ and $10\!:\!1$, $22\!:\!1$ and $21\!:\!1$, and in $41\!:\!1$ and $34\!:\!1$ for MURA Decoding and 3D-MLEM, respectively.

\subsection{Computation time}
\label{sec:res:comp_time}
To compare the computational performance of both reconstruction methods with the considered implementations, the average runtime for one image in a stack of images is presented here. The total runtime for reconstructing an image stack containing between 54 and 101 reconstructions with the 3D-MLEM algorithm for 40 iterations on a normal laptop computer with a 6-kernel Intel Core i7-9750H processor (2.6\,GHz) and 16\,GB of RAM ranged from $348$\,s to $579$\,s. Relative to the number of images per stack the average runtime is ($5.68 \pm 1.40$)\,s per image for $40$ iterations.

The runtime of MURA Decoding was obtained using a laptop equipped with a 12$^\text{th}$ generation Intel Core i7-12700H processor (2.3\,GHz) and 32\,GB of RAM. The total reconstruction time for image stacks containing between 60 and 240 reconstructions ranged from $31.2$\,ms to $547$\,ms with a mean of $(1.3 \pm 0.5$)\,ms per image. As a result, MURA Decoding is approximately $4{,}400$ times faster than 3D-MLEM with the considered implementations.

\section{Discussion}
\label{sec:discussion}
The aim of this paper was to systematically assess the axial resolution of a gamma camera equipped with a coded aperture collimator.
Generally, we found the preprocessing of the detector images to have a beneficial impact on the axial resolution, more for 3D-MLEM than for MURA Decoding, especially for distances below $40$\,mm. This is counter-intuitive, because one would think that Gaussian blurring should decrease the resolution since it acts as a low-pass filter and widens the peak.

\subsection{Reconstruction methods}
\label{sec:dis:reco}

Two improvements were made to the initial convolutional-based 3D-MLEM from~\cite{Mu2006}: first, a general normalization term is proposed that takes into account that photons emitted closer to the edge of the FOV are less likely to hit the detector. Second, transmission noise was added to the forward projection step. The absence of artifacts and the low background noise (see Fig.~\ref{fig:old_vs_new_3dmlem}) demonstrate the superiority of our proposed modifications.

Unlike MURA Decoding, the 3D-MLEM algorithm outputs a full 3D distribution since the contribution of all slices are taken into account.
This, however, requires more convolution operations. For each slice in each iteration the forward simulation, the backward simulation and another forward simulation with the updated slice are calculated. Additionally, the images that are processed - the entire detector image and the PSFs - are generally larger than for MURA Decoding. This makes it computationally expensive and thus slow, but more accurate.
Even though the runtime comparison was not carried out on the same computer and improvements in the implementation are likely possible, the order of magnitude that 3D-MLEM is slower than MURA Decoding is enough to say that it is too slow for intraoperative use.

Additionally, 3D-MLEM also provides the forward projection of the estimated reconstruction, which can be used for checking how well the reconstruction is in accordance with the acquired detector image (as an example see the image on the left in Fig.~\ref{fig:mlem_stuff}c).
Nevertheless, the long runtime renders the 3D-MLEM algorithm unsuitable for usage in an intraoperative scenario where the computation time substantially influences the practical utility of the gamma camera.

The ghost sources that the 3D-MLEM algorithm reconstructs for sources more than $40$\,mm away (see image on the right of Fig.~\ref{fig:mlem_stuff}c) are likely caused by the self-similarity of the mask pattern due to the $2\times2$ arrangement. 
From $40$\,mm on, more than the entire mask pattern is visible on the detector image. The surrounding margin (see the bottom of the detector images in Fig.~\ref{fig:mlem_stuff}b), which is not illuminated by the mask pattern, is relatively small and contains background noise, so the algorithm has only little area that would penalize a set of ghost sources that are partially contributing to the correct pattern in the center. 
It should be noted that these ghost sources do not completely prohibit a reconstruction, but they add an element of ambiguity.
This issue could also be addressed by adapting the algorithm to focus on the central portion of the detector image similar to MURA Decoding, which however would come with the cost of a narrower FOV.\\
The higher resolution in both lateral and axial directions makes 3D-MLEM interesting for large gamma cameras in SPECT systems, where runtime is less important.

\subsection{Comparing the axial resolution to literature}
\label{sec:dis:literature}

Setting the axial resolution in relation to the magnification factor $M$ from Eq.~(\ref{eq:M}), has the advantage of eliminating the dependency on the detector-to-mask distance $b$ and hence making a comparison with literature values easier. Fig.~\ref{fig:axial_res_plot} shows the axial resolution plotted against the magnification factor $M$. 

Unfortunately, only few values in literature exist with which we can compare our results: in a previous experiment~\cite{Russo2020} with the same coded aperture collimator but a slightly different detector-to-mask distance $b$, the same MURA Decoding as in this work was used. The authors evaluated the zone of best-focus of a ring-shaped object as the zone where the image contrast is maximum and constant within about 1\%. The zone of best-focus was reported as approximately $3$\,mm, while the lateral resolution was measured to be $0.6$\,mm at a source-to-mask distance of about $50$\,mm~\cite{Russo2020}. The given contrast profile is not equivalent to the CNR profile used in this work for assessing the axial resolution, but can still serve as a benchmark. When estimating the FWHM from the given graph, we obtain an FWHM of approximately $12$\,mm. This axial resolution is slightly better than the values reported in this paper but still provides good plausibility. 
Comparing the relative resolution as the ratio of axial to lateral resolution, we yield a ratio of $12\,\text{mm} / 0.6\,\text{mm} = 20$ at $50$\,mm distance. Our assessments of a ratio of $22$ and $21$ for MURA Decoding and 3D-MLEM at $50$\,mm distance are in good accordance with that. 

A rough classification for the reported 3D-MLEM results can be deducted from~\cite{Mu2006}: in the reported experiment, the authors placed sources shaped like an \quotes{H} and a \quotes{$>$}-symbol at $164$\,mm and $244$\,mm respective distances from the mask and reconstructed the scene with 3D-MLEM.
Since in their figure, the two sources appear separately in their corresponding planes, the axial resolution (FWHM of the CNR) must be smaller than half the distance between them, approximately $20$\,mm. Our values at comparable $M$ are lower than that and thus not contradictory. 

The presented findings of this work can answer the research question explicitly mentioned in~\cite{Fujii2012} on \quotes{What role iterative reconstruction algorithms [...] will play in improving Z resolution}: 
dividing the axial resolution obtained by the iterative 3D-MLEM algorithm by the axial resolution obtained by MURA-Decoding for each source position, gives an average factor of $0.3 \pm 0.1$. Hence, on average the axial resolution of 3D-MLEM is one third of the values from MURA Decoding, meaning with the first method we achieve a three times better axial resolution.

\subsection{Intraoperative application in SLNB}
\label{sec:dis:meaning}
In a recent review paper about intraoperative gamma cameras~\cite{Farnworth2023}, the authors state that for SLNB there is no consensus on the requirements for imaging parameter, including the lateral and axial resolution.
The authors of~\cite{Schellingerhout2002} consider a lateral resolution of $6$\,mm to be sufficient. However, SLNs can vary in size approximately between $5$ and $20$\,mm~\cite{Fujii2012, Kaissas2017}. Ideally, for a robust 3D localization of SLNs a spatial resolution much better than a few millimeters is required.
Nonetheless, the definition of an upper bound for the spatial resolution is needed before reliable assertions for the intraoperative application of compact gamma cameras in SLNB can be made. 

With the experimental setup presented here, an axial resolution below $5$\,mm can only be reached in source-to-mask distances lower than approximately $14$\,mm for MURA Decoding and lower than $50$\,mm for 3D-MLEM.
Therefore, the latter ensures a precise depth estimation for farther sources, at the cost of a long computational time that prevents the method to be used in real-time during intraoperative procedures. 
It must be emphasized, though, that the spatial resolution presented here was computed for point-like sources and consequently represents the best spatial resolution achievable. Further studies need to be pursued in order to assess how the axial resolution degrades when extended sources are being analyzed, as in the case of SLNs.

\subsection{Limitations}
\label{sec:dis:limits}

For pinhole and parallel collimators, measuring the spatial resolution as the FWHM of a point source is intuitive as the superposition principle allows the usage of the FWHM also when working with extended or multiple sources. 
However, the superposition assumption is more problematic in CAI. Extended sources are known to be reconstructed in lower quality than point sources~\cite{Mu2006, Fujii2012}. 
Furthermore, we do not have the entire 3D position of the sources, only the source-to-mask distance. Thus, an entire 3D localization error as in \cite{Kaissas2017}, cannot be presented here. 
Nor did we test different values for detector-to-mask distance $b$.

In an intraoperative experiment with pigs, a reduction of the dynamic range was observed when a very bright source was present. It was called \quotes{concentration effect} and makes weaker sources less visible~\cite{Fujii2012}. The 3D-MLEM algorithm is supposedly immune to this effect, but that has not been specifically investigated~\cite{Mu2006}.
However, with a CdTe photon counting detector of the Medipix2/Timepix2 series~\cite{Russo2020}, we observed that a $^\text{241}$Am gamma source with an activity of $1\,\mu$Ci is still visible when a $1$\,mCi $^\text{241}$Am source is placed nearby and used as a background. This is possible thanks to the extended counting linearity range, the practical immunity to noise and the pixel-wise functioning of our photon counting detector, with respect to scintillator-based, Anger-logic based gamma cameras.\\
Another aspect that was outside the scope of this paper is the impact of a lateral shift of the source: due to near field artifacts a degradation in lateral resolution has been reported when a source is off-center~\cite{Accorsi2001a}. A decrease in axial resolution can be expected, but the exact effect remains an open question. 

A thin mask causes small collimation artifacts but brings a high transmission rate. That means septal penetration, which in turn widens the PSF and makes the shadow more fuzzy. The question on how transmission affects the lateral and axial resolution remains open.

\section{Conclusion \& Outlook}
\label{sec:cando}
In this paper, we systematically assessed the axial resolution of a gamma camera equipped with a coded aperture collimator with $0.08$\,mm holes. We calculated the FWHM by reconstructing images closely around the true source distance and subsequently fitting a Gaussian curve with offset to the extracted CNR profile. 
To increase reproducibility, the CNR profile along the z-direction of the obtained image stack was calculated by a semi-automatic algorithm.
In addition to the most commonly used reconstruction method -~MURA Decoding~- a 3D-MLEM algorithm has been adapted to deal with transmission noise and a more general camera setup.\\

\noindent This work completes our understanding of the spatial resolution in all three dimensions of our experimental gamma camera. 
Our presented findings show that CAI makes the depth estimation of single-point sources in the near field possible. How precisely the source-to-mask distance can be estimated depends on the reconstruction method. In the two different reconstruction methods we compared, a large difference in their axial resolution was observed. MURA Decoding was found to be fast and with sufficient precision for the nuclear medicine imaging task, while the 3D-MLEM algorithm reconstructs with higher precision, both in lateral and axial directions, but it is much slower than MURA Decoding: this might be a limitation for real-time reconstruction with a compact gamma camera.

Even though the axial resolution was 10 to 40 times worse than the lateral resolution -~for both reconstruction methods~-, gaining depth information from a single image capture makes CAI a unique collimation technique.
When operating in the near field of the camera, obtaining a 3D position of point sources adds to the benefits of a coded aperture, apart from the higher resolution and photon harvesting. \\

\noindent Determining the localization error was beyond the scope of this work. Now that we have concrete figures for the axial resolution, further investigations are underway to assess the full 3D position of point sources, comparing the true source position with its estimate based on a single detector image.
In order to further analyze the potential intraoperative use for a gamma camera with coded aperture collimator, experiments closer to the real-world use cases are essential. These experiments require a new mask: since the most commonly used radionuclide in nuclear medicine is $^{\text{99m}}$Tc~\cite{Peterson2011}, whose main emission is at $140.5$\,keV, a thicker mask is required to prevent transmission. Moreover, taking into account the standard diameter of SLNs, a wider FOV is required for the use of CAI in this field, as the source's occupation of the FOV must be limited to obtain an image with sufficient contrast. For these reasons, a new coded aperture collimator with $0.25$\,mm holes on a $1$\,mm thick Tungsten sheet is currently under development by the University \& INFN Naples group. 
Furthermore, more research might be aimed at accelerating the 3D-MLEM algorithm or reducing the number of necessary iterations to make it capable of running in real-time. Only that would render it feasible for intraoperative usage.
Additionally, CAI is known for its problems with extended sources and thus its influence on 3D localization must be further investigated.

\subsection*{Supplementary information}

\subsection*{List of abbreviations:}
ASIC: application-specific integrated circuit,
CAI: coded aperture imaging, 
CNR: contrast-to-noise ratio, 
FWHM: full width at half maximum, 
FOV: field of view, 
MLEM: maximum likelihood expectation maximization, 
MURA: modified uniformly redundant array, 
NTHT: no-two-holes-touching, 
PSF: point-spread function, 
ROI: region of interest, 
SPECT: single photon emission tomography,
SLN: sentinel lymph node, 
SLNB: sentinel lymph node biopsy, 
THT: two-holes-touching,
URA: uniformly redundant array.

\subsection*{Declarations}
\subsubsection*{Ethics approval and consent to participate}
Not applicable.

\subsubsection*{Consent for publication}
Not applicable.

\subsubsection*{Availability of data and material}
The dataset generated and analyzed during the current study are available in the Zenodo repository at \url{https://zenodo.org/doi/10.5281/zenodo.8315861}.

\subsubsection*{Competing interests}
The authors declare that they have no competing interests.

\subsubsection*{Funding}
This project was partly funded by Zentrales Innovationsprogramm Mittelstand (ZIM) under grant KK5044701BS0 from the German Federal Ministry for Economic Affairs and Climate Action. Open access funding was provided by the library of the Karlsruhe Institute of Technology. An additional part of the funding was provided by the INFN Naples within the Medipix4 project. 

\subsubsection*{Authors' contributions}
Contributions are listed according to the CRediT system. 
TM: conceptualization, formal analysis, investigation, methodology, software, validation, visualization, and writing (original draft). 
LAC: conceptualization, formal analysis, investigation, methodology, software, validation, and writing (original draft). 
PR: conceptualization, project administration, resources, supervision, and writing (review and editing). 
WN: conceptualization, funding acquisition, supervision, and writing (review and editing).
JH: conceptualization, funding acquisition, project administration, resources, supervision, and writing (review and editing). 
All authors read and approved the final manuscript.

\subsubsection*{Acknowledgements}
The authors would like to acknowledge the MEDIPIX collaboration (\url{https://medipix.web.cern.ch/home}) and the INFN which, as a  member of Medipix2 and Medipix4 Collaborations at CERN, granted use of the MinipixEDU Timepix3 detector used in this work. The MURA decoding algorithm was written and provided to the Napoli team by Dr. R. Accorsi (formerly at MIT, Cambridge, MA, USA). We, additionally, acknowledge the data storage service SDS@hd supported by the Ministry of Science, Research and the Arts Baden-W{\"u}rttemberg (MWK) and the German Research Foundation (DFG) through grant INST 35/1314-1 FUGG and INST 35/1503-1 FUGG and support by the KIT-Publication Fund of the Karlsruhe Institute of Technology.

\backmatter

\FloatBarrier
\begin{appendices}

\section{Additional 3D reconstructions}
\label{app:50and100}

\begin{figure}[ht]
    \centering
    \begin{subfigure}[b]{1\textwidth}
        \centering
        \includegraphics[width=1\textwidth, trim=0 5 10 5, clip]{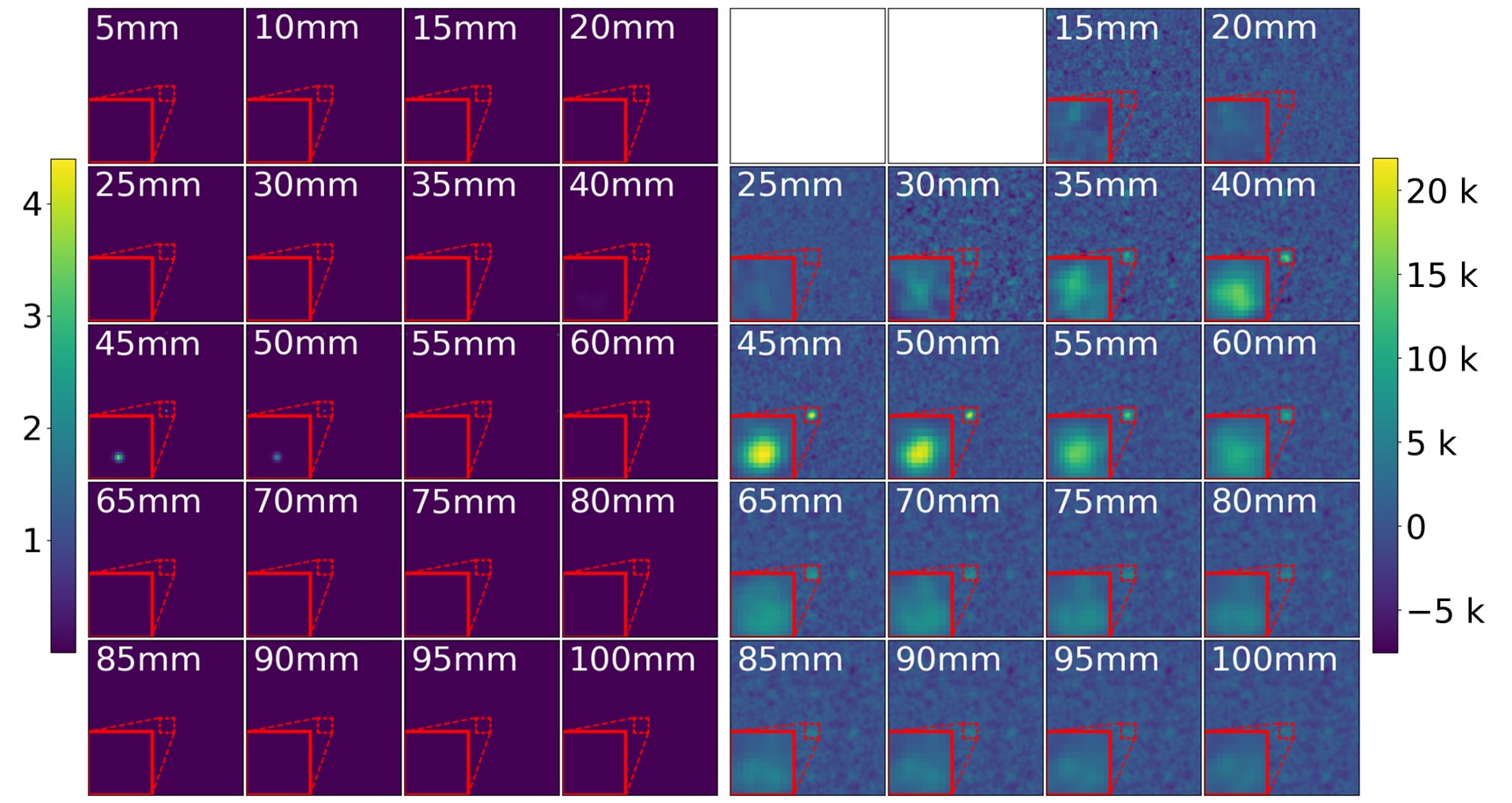}
        \caption{}    
        \label{fig:3d_comp_50mm}
    \end{subfigure}
    \begin{subfigure}[b]{1\textwidth}
        \centering
        \includegraphics[width=1\textwidth, trim=0 5 15 5, clip]{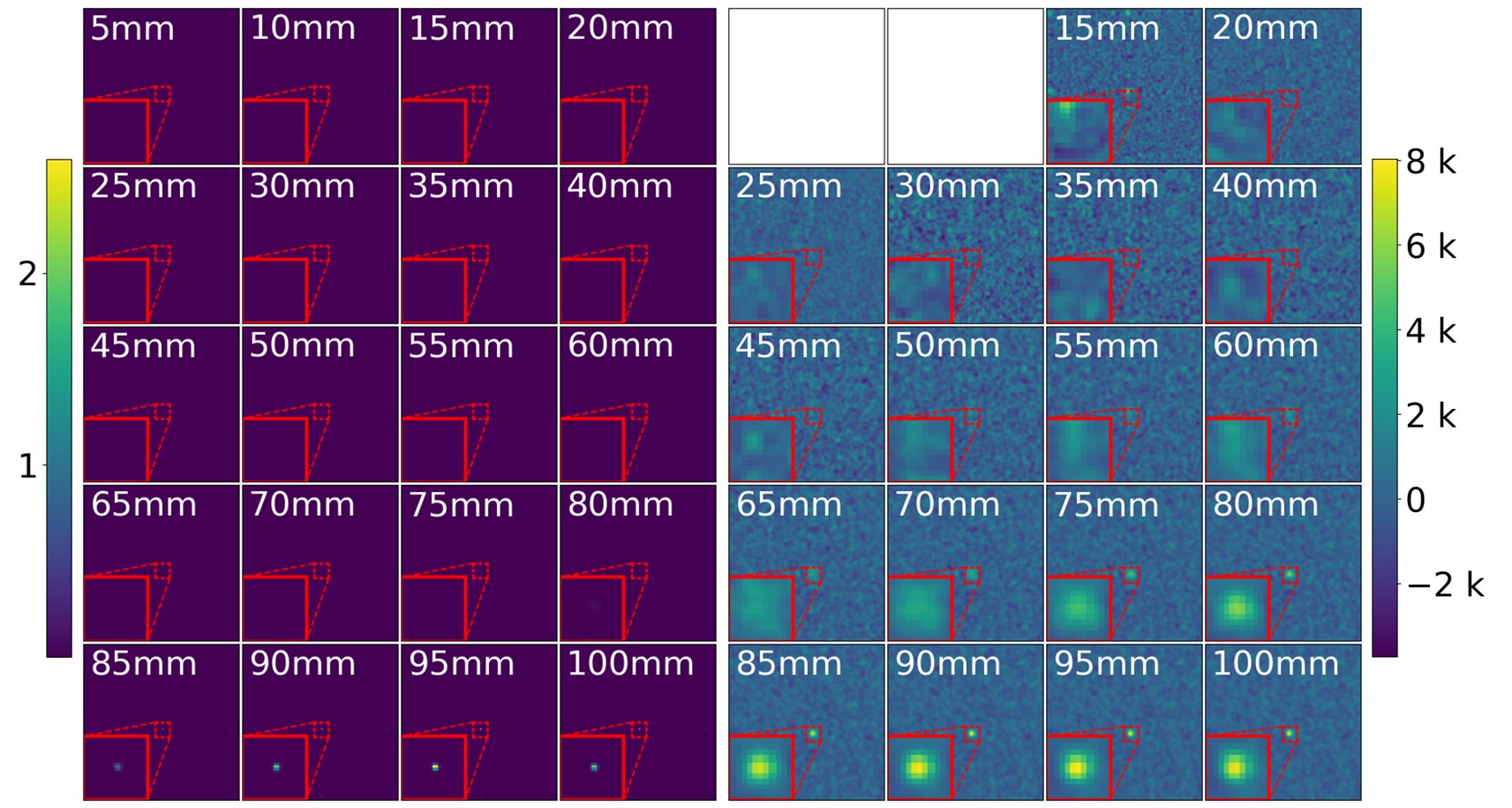}
        \caption{}    
        \label{fig:3d_comp_100mm}
    \end{subfigure}
    \caption{The entire image stack of reconstructing the source at 20 equidistant image planes from $5$ to $100$\,mm with MURA Decoding (right) and 3D-MLEM (left) for the source at $50$\,mm (\subref{fig:3d_comp_50mm}) and $100$\,mm (\subref{fig:3d_comp_100mm}). The area marked by the red square has been magnified for better visualization.}
    \label{fig:3d_comp_50and100}
\end{figure}
\FloatBarrier

\section{Reconstruction from raw and preprocessed detector image}
\label{app:raw_vs_prep_z20@20}

\begin{figure}[ht]
\centering
\includegraphics[width=0.7\textwidth, trim=5 5 5 5, clip]{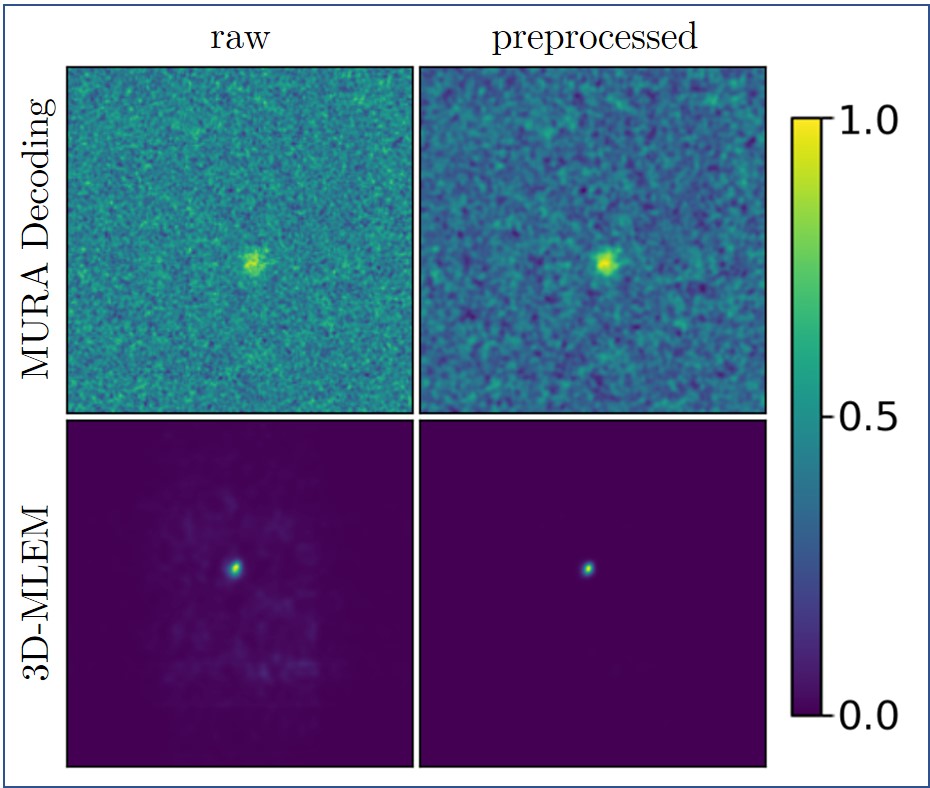}
\caption{All images show the reconstruction of the raw (left) and preprocessed (right) detector image of the source at source-to-mask distance of $20$\,mm at the $20$\,mm plane. The top row shows the reconstructions from MURA Decoding while the bottom row shows the 3D-MLEM results. Notice the higher background noise in the reconstructions from the raw detector image. For a better comparison all images were normalized to the pixel intensity range of $0$ to $1$.}
\label{fig:recos_prep_vs_raw}
\end{figure}

\FloatBarrier
\end{appendices}

\bibliography{references}

\end{document}